## 1.6 International Conferences of Bibliometrics

Grischa Fraumann, Rogério Mugnaini, and Elías Sanz-Casado

**Abstract:** Conferences are deeply connected to research fields, in this case bibliometrics. As such, they are a venue to present and discuss current and innovative research, and play an important role for the scholarly community. In this article, we provide an overview on the history of conferences in bibliometrics. We conduct an analysis to list the most prominent conferences that were announced in the newsletter by ISSI, the International Society for Scientometrics and Informetrics. Furthermore, we describe how conferences are connected to learned societies and journals. Finally, we provide an outlook on how conferences might change in future.

### Introduction

Conferences are deeply connected to research fields, in this case bibliometrics. As such, they are a venue to present and discuss current and innovative research, and play an important role for the scholarly community. They also serve as a venue to play a variety of roles, strengthening the so-called invisible colleges (Zuccala, 2006), and are important for gaining scientific reputation (Söderqvist and Silverstein, 1994). Furthermore, conference awards and committee memberships are a marker of prestige among scholars (Jeong, Lee and Kim, 2009). This chapter provides an overview on international conferences in bibliometrics, and what role they play in the history and institutionalization of bibliometrics. Proceedings papers are published in conference proceedings, and such proceedings are also indexed, for example, since 1990 by the *Conference Proceedings Citation Index* (*CPCI*) (Sugimoto and Larivière, 2018) and in *Scopus* (Gingras, 2016). Apart from books and journal articles, proceedings papers have a long tradition in disseminating research (Sugimoto and

**Grischa Fraumann**, Research Assistant at the TIB Leibniz Information Centre for Science and Technology in the R&D Department, PhD Fellow at the University of Copenhagen in the Department of Communication, Research Affiliate at the "CiMetrias: Research Group on Science and Technology Metrics" at the University of São Paulo (USP), gfr@hum.ku.dk
**Rogério Mugnaini**, Professor of Library and Information Science at the University of São Paulo (USP), where he leads the research group "CiMetrias: Research Group on Science and Technology Metrics", mugnaini@usp.br
**Elías Sanz-Casado**, Full Professor in the Department of Library and Information Science at the Carlos III University of Madrid, Director of the research group "Laboratory for Metric Information Studies" (LEMI), leads the "Research Institute for Higher Education and Science" (INAECU) which is made up of members of Carlos III University of Madrid and Autonomous University of Madrid, elias@bib.uc3 m.es

Larivière, 2018), and are used as datasets for bibliometric (Glänzel et al., 2006; Lisée, Larivière, and Archambault, 2008) and altmetric studies (Thelwall, 2019). On the one hand, not all conference proceedings are indexed, which makes such citation indexes sometimes incomplete (Sugimoto and Larivière, 2018). On the other hand, some conferences publish their proceedings as journal special issues or books, which can lead to the indexing of proceedings papers. Proceedings papers play a particular role in natural sciences and medicine (Ball, 2017). They represent an important means of publication in computer science (Fathalla et al. 2018), while they might be rather irrelevant in other disciplines, such as sociology (Jeong, Lee and Kim, 2009). As such, they also contribute to faster knowledge sharing than journal articles.

## A Historical Sketch on Conferences in Bibliometrics

There are several examples of early conferences in bibliometrics. In 1946, two international conferences on scientific information were the first events, organised by the Royal Society of London (Gingras, 2016). The goal was to develop new forms of indexing scientific literature. This was also related to the exponential amount of scientific literature that led to the foundation of a citation index as part of Garfield's Institute for Scientific Information (ISI) in 1963. The first international conference was held in 1974, titled "Toward a Metric of Science: The Advent of the Science Indicators" (Gingras, 2016). 1987 is considered as the start of a new era in international conferences, since a predecessor of the conferences organised by the International Society for Scientometrics and Informetrics (ISSI) was held for the first time.

This led in 1993 to the foundation of a learned society, namely ISSI (Gingras, 2016). Conferences, journals, and learned societies go in line with a consolidation of an academic discipline, as the foundation of the journal *Scientometrics* in 1978 demonstrates (Gingras, 2016). Most conferences nowadays also experience other forms of scholarly communication, such as the live tweeting about conference presentations and discussions (Holmberg, 2015). Such tweets may also lead to collaborations for researchers that do not physically attend a conference (Holmberg, 2015). Furthermore, conference reports are often also communicated via blogs. Other forms of communication are live streams or archived videos of conference presentations that are available on dedicated online platforms (Plank et al., 2019).

## An Overview of International Conferences and its Relation to Learned Societies

ISSI is one of the largest learned societies in bibliometrics, among others. Established in 1993 by a group of researchers during a conference in Berlin, the Netherlands-based association coordinates the ISSI Conference, a members' directory, a blog, and a quarterly newsletter. Conference participants and society members get access to the ISSI conference proceedings, which contain all proceedings papers. ISSI also advocates for several international initiatives, such as the Initiative on Open Citations (I4OC), which has been supported by ISSI in an open letter (Sugimoto, Murray, and Larivière, 2018).

The influence by the society on the research fields can be observed by its significant number of members and conference participants. The conferences used to run independently (ISSI, 2019), and in 2019 were held for the first time together with the European Network of Indicators Designers (ENID). The biennial conference is held in different locations around the world, and the abovementioned first conference in 1987 was called "International Conference on Bibliometrics and Theoretical Aspects of Information Retrieval". As happens often within the scholarly community, anecdotal evidence suggests that the conference was started by a discussion between two researchers, and the question "Shouldn't we start a biennial international conference on informetrics?" (ISSI, 2015). Since 1993 it bears the same name as of today (Hood and Wilson, 2001; ISSI, 2019). The conference is reported to be one of the world's largest and most prestigious conferences (Gorraiz et al., 2014).

Considering the influence of ISSI, the quarterly newsletter is used as a dataset to query past and ongoing international conferences in this research field. A related data collection method has been carried out, for instance, by Söderqvist and Silverstein (1994) for conferences in immunology and Jeong et al. (2009) for conferences in bioinformatics. The *ISSI Quarterly Newsletter* started in 2005, and published 59 issues until September 2019, as of October 28, 2019. All newsletters are publicly available also for non-members. The newsletter in PDF has an ISSN and is curated by 10 members of an editorial board according to guidelines and potential authors need to submit proposals. This is to say the structure is rather similar to a magazine than to a newsletter. Apart from discussions on current research and the introduction of members as well as other news of the society, the newsletter includes conference announcements, call for papers and conference reports. Formats such as workshops, meetings, symposia, forums, summer schools, PhD courses, seminars, and other training courses were excluded from the selection in this article. All quarterly newsletters since the start were downloaded. This approach might have some limitations. For example, certain conferences might be excluded, because they were announced somewhere else. Still, the approach provides a glimpse into the world of conferences on bibliometrics. The conferences were ordered according to the frequency or status, and additional information on their listing in *Conference Proceedings Citation Index (CPCI)* as well as from conference websites was provided if available (see Table 1). The conferences that were discontinued over time are not included in Table 1, except for the UK Social Networks Conference that was merged with two other conferences.

**Table 1:** International conferences in bibliometrics (*N*=11) ordered by frequency/ status including information on their listing in *CPCI* (source: *ISSI Quarterly Newsletter* March 2005 until September 2019, number 1–59, retrieved from http://issi-society.org/publications/issi-newsletter/ [July 15, 2020]).

| Conference name | Frequency/ Status | Listed in CPCI | Description |
| --- | --- | --- | --- |
| International Conference "Impact of Science" – Measuring and Demonstrating the Societal Impact of Science | Several times per year | - | "[A] conference […] to discuss measuring and demonstrating the societal impact of science" (https://scienceworks.nl/impact-of-science-2015/ [July 15, 2020]) |
| Triple Helix Conference | Annually | + | "The Triple Helix model presents an opportunity to achieve innovation outcomes for the socio-economic good through collaboration with multi-stakeholders within academia, industry and government spheres." (https://triple-helix.co.za/ [July 15, 2020]) |
| S&T Indicators Conference | Annually | + | Conference on Science and Technology Indicators |
| InSciT Conference | Annually | - | Conference on Science of Team Science |
| WissKom conference | Annually | - | Conference of the Central Library at Forschungszentrum Jülich, Germany |
| CARMA: International Conference on Advanced Research Methods and Analytics | Annually | + | "Research methods in economics and social sciences are evolving with the increasing availability of Internet and Big Data sources of information. As these sources, methods, and applications become more interdisciplinary, the […] International Conference on Advanced Research Methods and Analytics (CARMA) aims to become a forum for researchers and practitioners to exchange ideas and advances on how emerging research methods and sources are applied to different fields of social sciences as well as to discuss current and future challenges." (http://www.carmaconf.org/ [July 15, 2020]) |
| iConference | Annually | + | "[…] insights on critical information issues in contemporary society." (https://ischools.org/iConference [July 15, 2020]) |

**Table 1** *(Continued)*

| Conference name | Frequency/ Status | Listed in CPCI | Description |
|---|---|---|---|
| International Conference on Webometrics, Informetrics and Scientometrics (WIS) & COLLNET Meeting | Annually | + | "[…] all aspects of webometrics, informetrics and scientometrics." (http://collnet2019.dlut.edu.cn/meeting/index_en.asp?id=2676 [July 15, 2020]) |
| International Conference on Scientometrics and Informetrics (ISSI) | Biennial | + | "The goal of [the] ISSI [conference] is to bring together scholars and practitioners in the area of informetrics, bibliometrics, scientometrics, webometrics and altmetrics to discuss new research directions, methods and theories, and to highlight the best research in this area." (https://www.issi2019.org/ [July 15, 2020]) |
| Atlanta Conference on Science and Innovation Policy | Biennial | + | "The Atlanta Conference on Science and Innovation Policy provides a showcase for the highest quality scholarship from around the world addressing the challenges and characteristics of science and innovation policy and processes." (http://www.atlconf.org/ [July 15, 2020]) |
| UK Social Networks Conference | Discontinued | - | The UK Social Networks Conference merged with the Applications of Social Network Analysis to form the European Conference on Social Networks (EUSN) (https://www.eusn2019.ethz.ch/ [July 15, 2020]) |

To the best of the authors' knowledge, there is no comprehensive discipline-specific database available that includes all conferences in bibliometrics, while there are several databases worldwide, for example discipline-specific ones, such as *dblp computer science bibliography* (Ley, 2002). Generally speaking, these databases list proceedings papers that are linked to, for example, authors and conferences. Apart from the conferences mentioned in Table 1, one could add the Altmetrics Conference, an annual conference that provides a venue for research and other initiatives on altmetrics, that is metrics to track research articles online (Priem et al., 2010), but only the related Altmetrics Workshop was mentioned in the *ISSI Quarterly Newsletter*, and workshops were excluded from this selection. Other available databases on a national level might include the Brazilian *Lattes Platform* (Marques 2015) that shows CVs of researchers and their publications as well as the attended national and international conferences (Mugnaini et al., 2019). A similar database might also be pro-vided by institutional, national or international CRIS (Current Research Information Systems) (Sivertsen, 2019), such as *DSpace-CRIS* (Palmer et al., 2014) or *VIVO* (Börner et al., 2012; Conlon et al., 2019). Nevertheless, there are certain ongoing initiatives that might provide in future a more nuanced view on international conferences. One example is the *ConfIDent platform* that is developed as part of a research project funded by the German Research Foundation (DFG). The project objective is to develop a Wiki-based platform that takes into account the needs of scholarly communities, and offers a curated list of conferences. By structuring the conference data according to requirements of interoperability on the technical side and to academic demands on the social side, the system aims to present the possibility for a sociotechnical quality assessment of the content (Hagemann-Wilholt, Plank, and Hauschke, 2019; Hagemann-Wilholt, 2019; Sens and Lange, 2019). Generally speaking, proceedings papers may also be linked to ORCID IDs of researchers, that is personal identifiers (Dreyer et al., 2019). There are also other existing online platforms, such as *Open Research* (Fathalla et al., 2019) and *ConfRef*. Additionally, there are further initiatives to develop persistent identifiers (PIDs) for conferences (Crossref, 2019). There are also initiatives underway to develop a semantic representation of scientific events (Fathalla and Lange, 2018) in knowledge graphs, and to make structured queries available to the wider public (Fathalla, Lange, and Auer, 2019a). Such datasets may also be used to rank conferences or explore the impact of these events (Fathalla, Lange and Auer, 2019b; Hansen and Budtz Pedersen, 2018; Hauschke, Cartellieri, and Heller, 2018; Altemeier, 2019).

## Conclusions

Conferences are important vehicles to disseminate research and to network with peers. Conferences aligned with journals and learned societies serve an important role in the institutionalization of bibliometrics. Bibliometricians and altmetricians also develop metrics based on data from conferences. There are ongoing initiatives to develop persistent identifiers for conferences. How will the future of conferences look like? Live streaming and other online tools, among others, might have changed the importance of traveling to conferences, and social media makes it possible to join the discussion remotely, but conferences will most probably remain prominent in the academic world, because the social component of meeting other researchers is one of the most important aspects. However, the climate crisis and coronavirus pandemic will most probably require fundamental changes for most academic conferences, and will accelerate digital alternatives and make it less necessary to physically attend these events (Viglione, 2020).

**Acknowledgements**
This chapter was funded by the German Federal Ministry of Education and Research (BMBF)
under grant numbers 01PU17019 (ROSI – Reference Implementation for Open Scientometric Indicators)
and 16PGF0287 (BMBF Post-Grant-Fund).

**After-publication Corrigendum**

p. 67:
As happens often within the scholarly community, the conference was started by a discussion between two researchers, Leo Egghe and Ronald Rousseau, and the question "Shouldn't we start a biennial international conference on informetrics?" (ISSI, 2015).